\documentclass[twocolumn]{article}
\usepackage[utf8]{inputenc}
\usepackage{pdfpages}
\usepackage{subfigure}
\usepackage{authblk}
\usepackage{multirow}
\usepackage{graphics}
\usepackage{float}
\usepackage{amsmath}
\usepackage{url}
\usepackage{booktabs}
\usepackage{gensymb}

\title{Applying VertexShuffle Toward 360-Degree Video Super-Resolution on Focused-Icosahedral-Mesh}

\author{Na Li}
\author{Yao Liu}
\affil{Department of Computer Science, Binghamton University}

\date{}

\begin{document}

\maketitle

\begin{abstract}

With the emerging of 360-degree image/video, augmented reality (AR) and virtual reality (VR), the demand for analysing and processing spherical signals get tremendous increase. However, plenty of effort paid on planar signals that projected from spherical signals, which leading to some problems, e.g. waste of pixels, distortion.
Recent advances in spherical CNN have opened up the possibility of directly analysing spherical signals. However, they pay attention to the full mesh which makes it infeasible to deal with situations in real-world application due to the extremely large bandwidth requirement.
To address the bandwidth waste problem associated with 360-degree video streaming and save computation, we exploit Focused Icosahedral Mesh to represent a small area and construct matrices to 
rotate spherical content to the focused mesh area.
We also proposed a novel VertexShuffle operation that can significantly improve both the performance and the efficiency compared to the original MeshConv Transpose operation introduced in UGSCNN~\cite{jiang2019spherical}.
We further apply our proposed methods on super resolution model, which is the first to propose a spherical super-resolution model that directly operates on a mesh representation of spherical pixels of 360-degree data.
To evaluate our model, we also collect a set of high-resolution 360-degree videos to generate a spherical image dataset.
Our experiments indicate that our proposed spherical super-resolution model achieves significant benefits in terms of both performance and inference time compared to the baseline spherical super-resolution model that uses the simple MeshConv Transpose operation.
In summary, our model achieves great super-resolution performance on 360-degree inputs, achieving 32.79 dB PSNR on average when super-resoluting 16x vertices on the mesh.

\end{abstract}

\section{Introduction}

360-degree image/video, also known as spherical image/video, is an emerging format of media that captures views from all directions surrounding the camera. 
Unlike traditional 2D image/video that limits the user's view to wherever the camera is facing during capturing, a 360-degree image/video allows the viewer to freely navigate a full omnidirectional scene around the camera position.

Despite its substantial promise of immersiveness, the utility of streaming 360-degree video is limited by the huge bandwidths required by most streaming implementations.
When watching a 360-degree video, users can only watch a small portion of the full omnidirectional view. That is, while the 360-degree video encodes frames that cover the full $360\degree \times 180\degree$ field-of-view (FoV), the user may only observe a ``view'' of $100\degree \times 100\degree$ FoV of the omnidirectional frame at a time.
If the omnidirectional frame is projected to the 2D frame using the equirectangular projection~\cite{equirectangular}, then only roughly 15\% of the pixels of the frame is viewed. The rest 85\% pixels are not viewed, and are thus wasted. 

To allow the users to observe ``views'' in high enough quality, 
full omnidirectional frames must be transmitted at 4K or 8K resolution.
Streaming videos at 4K or 8K resolution requires a significant amount of network bandwidth (e.g., 100 Mbps for 8K video streaming) that may not be supported by most users' network connections.  

To address the bandwidth-waste problem, we proposed a efficient mesh representation, Focused Icosahedral Mesh, supports our model to focus on the more interesting portion of a sphere instead of the full mesh. It is more flexible and efficient.

\begin{figure}[t]
  \centering
  \vspace{3em}
  \includegraphics[width=\columnwidth]{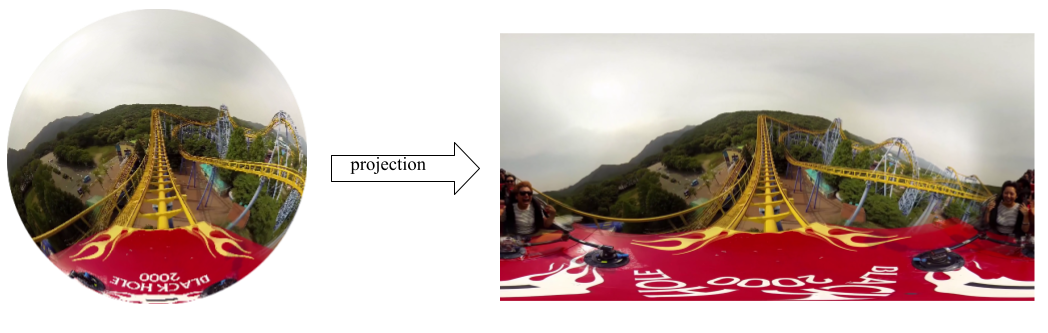}
   \caption{This left shows the same 360-degree content as pixels on a sphere, which is the most natural way for representing 360-degree data. The right shows a 360-degree image encoded in the equirectangular projection, which is a widely used spherical projection for representing 360-degree images. However, projecting spherical signals to the 2D plane introduces distortion, e.g., the north and south pole areas.\protect\footnotemark}
\label{fig:introduction}
\end{figure}

\footnotetext{The original image is from the following video in the 360-degree video head movement dataset~\cite{corbillon2017360}: \protect\url{https://www.youtube.com/watch?v=8lsB-P8nGSM}}

Another problem is that the omnidirectional views captured by 360-degree cameras are most naturally represented as uniformly dense pixels over the surface of a sphere (as shown in Figure \ref{fig:introduction} (left)).
When spherical pixels are projected to planar surfaces, distortions are introduced. 
For example, the equirectangular projection~\cite{equirectangular} is a widely used spherical projection for representing 360-degree data. 
However, significant distortions can be observed around the north and south pole areas, as shown in Figure \ref{fig:introduction} (right). 

Such distortions can reduce the efficiency of CNN operations by adding ``over-represented'' pixels, e.g., the regions near the north and south poles in the equirectangular projection. 
Further, training a CNN directly on the distorted representation could cause CNN models to learn characteristics of the planar distortion rather than relevant details of the high resolution representation.

Recent work~\cite{jiang2019spherical,cohen2017convolutional} pay attention to play Convolutional operation directly on spherical signals to prevent the distortion problem. Their works show that it is possible to analyze spherical signals directly without 2D projections. 
Furthermore, extensive experiments were conducted in these works to show the efficiency of their proposed spherical CNNs.

Motivated by 2D PixelShuffle~\cite{shi2016real}, we also proposed our VertexShuffle that achieves great performance and parameter efficiency on mesh representation, which improved a lot based on MeshConv Transpose proposed in UGSCNN~\cite{jiang2019spherical}.

To illustrate the efficiency of our proposed Focused Icosahedral Mesh representation and VertexShuffle, we apply our methods on a popular problem in computer vision, Super-Resolution~\cite{irani1991improving}, which aims at recovering high-resolution images and videos from low-resolution images and videos. 

In this paper, inspired by recent advances in spherical CNNs~\cite{jiang2019spherical,cohen2017convolutional} and state-of-the-art 2D super-resolution methods~\cite{dong2014learning, dong2015image, dong2016accelerating, zhang2018residual, tai2017image}, We proposed an efficient Focused Icosahedral Mesh representation to better utilize the computational resources and a novel VertexShuffle operation that can significantly improve both the performance and the efficiency compared to the original MeshConv Transpose operation introduced in UGSCNN~\cite{jiang2019spherical}. For evaluation, due to the lack of former spherical super resolution dataset, we also created a spherical super-resolution dataset from ten 360-degree videos in high resolution.

In summary, our paper makes the following main contributions:

\begin{itemize}
    \item We create a Focused Icosahedral Mesh representation of the sphere to efficiently represent spherical data, which not only saves computational resources but also improves memory efficiency.
    \item We create a novel VertexShuffle operation, inspired by the 2D PixelShuffle~\cite{shi2016real} operation. The Vertex operation significantly increases both the visual quality metric (peak signal-to-noise ratio (PSNR)) and inference time over comparable transposed convolution operations.
    \item We are the first to propose a super-resolution model that directly operates on a mesh representation of spherical pixels of 360-degree data. 
    \item We create a 360-degree super-resolution dataset from a set of high resolution 360-degree videos for evaluation. 
    \item Results show that our proposed SSR model achieves great super-resolution performance on 360-degree inputs, achieving 32.79 dB PSNR on average when super-resoluting 16x vertices on the mesh.
\end{itemize}

\section{Related Work}
\subsection{360-degree video}
Despite its potential for delivering more-immersive viewing experiences than standard video streams, current 360-degree video implementations requires bandwidths that are too high to deliver adequate experiences for many users.
 
Numerous approaches have been proposed for improving 360-degree bandwidth efficiency. These approaches have both attempted to improve the efficiency of how the 360-degree view is represented during transmission~\cite{petrangeli2017http,xie2017360probdash,nasrabadi2017adaptive,zare2016hevc,corbillon2017viewport,graf2017towards,sun2018multi,mahzari2018fov,qian2018flare,guan2019pano} as well as improving a system's ability to avoid delivering unviewed pixels~\cite{zhou2017measurement}.
Only recently have super-resolution (SR) approaches been proposed in conjunction with 360-degree video delivery~\cite{dasaristreaming,chen2020sr360}.

To avoid distortion problem in projecting 360-degree video to 2D planes, Xiong et al.~\cite{xiong2018snap} developed a reinforcement learning approach to select a sequence of rotation angles to minimize interest area near or on the cube boundaries. 

Marc et al.~\cite{eder2020tangent} proposed a spherical image representation that mitigates spherical distortion by rendering tangent icosahedron faces to a a set of oriented, low-distortion images to icosahedron faces. They also presented utilities of applying standard CNN on spherical data.

\subsection{Spherical convolutional neural networks}
Spherical CNN has been studied by the computer vision community recently as a number of real-world applications require processing signals in the spherical domain, including self-driving cars, panoramic videos, omnidirectional RGBD images, and climate science. 

Recently works such as Cohen et al.~\cite{cohen2017convolutional} gave theoretical support of spherical CNNs for rotation-invariant learning problems, which is important for problems where orientation is crucial to the model performance. 
They first introduced the concepts of $S^2$ and $SO(3)$. $S^2$ can be defined as the set of points on a unit sphere, and $SO(3)$ is the rotation group in Euclidean three dimensional space. 

They replaced planar correlation with spherical correlation, which can be understood as the value of output feature map evaluated at rotation $R \in SO(3)$ computed as an inner product between the input feature map and a filter, rotated by $R$. Furthermore, they implemented the generalized Fourier transform for $S^2$ and $SO(3)$.

Later, Cohen et al.~\cite{cohen2019gauge} introduced a theory that is equi-variance to symmetry transformations on manifolds. They further prompt a gauge equi-variant CNN for signals on the icosahedron using the icosahedral CNN, which implements gauge equi-variant convolution using a single conv2d call, making it a highly scalable and practical alternative to spherical CNNs. 

UGSCNN~\cite{jiang2019spherical} is another recent work in spherical CNN. It presents a novel CNN approach on unstructured grids using parameterized differential operators for spherical signals. They introduce a basic convolution operation, called MeshConv, that can be applied on meshes rather than planar images. 
It achieves significantly better performance and parameter efficiency compared to state-of-the-art network architectures for 3D classification tasks since it does not require large amounts of geodestic computations and interpolations. 

Zhang et al.~\cite{zhang2019orientation} proposed to perform semantic segmentation on omnidirectional images by designing an orientation-aware CNN framework for the icosahedron mesh. They introduced fast interpolation of kernel convolutions and presented weight transfer from learned through classical CNNs to perspective data.
Recently, Eder et al.~\cite{eder2020tangent} proposed a spherical image representation that mitigates spherical distortion by rendering a set of oriented, low-distortion images tangent to icosahedron faces. They also presented utilities of their approaches by applying standard CNN to the spherical data.

While these existing works demonstrate their effectiveness in classification and segmentation tasks,  the super-resolution task was not considered. 
In this work, we found it possible to apply their work on the super-resolution task. Our work is based on the proposed MeshConv operation since it achieves better performance and parameter efficiency than other spherical convolutional networks. We also conduct experiment to show significant improvements over the baseline spherical super-resolution model that uses the simple MeshConv Transpose operation.

\subsection{Super resolution}
The super-resolution field has advanced rapidly from its origins in the deep learning age.
The SRCNN~\cite{dong2014learning, dong2015image} model was the first to apply CNNs to SR.
FSRCNN~\cite{dong2016accelerating} was an evolution of SRCNN. It operated directly on a low-resolution input image and applied a a deconvolution layer to generate the high-resolution output. 
VDSR~\cite{kim2016accurate} was the first to apply residual layers~\cite{he2016deep} to the SR task, allowing for deeper SR networks. 
DRCN~\cite{kim2016deeply} introduced recursive learning in a very deep network for parameter sharing. 
Shi et al.~\cite{shi2016real} proposed ``PixelShuffle'', a method for mapping values at low-resolution positions directly to positions in a higher-resolution image more efficiently than the deconvolution operation.
SRResNet~\cite{ledig2017photo} introduced a modified residual layer tailored for the SR application. 
EDSR~\cite{lim2017enhanced} further modified the SR-specific residual layer from SSResNet and introduced a multi-task objective in MDSR.
SRGAN~\cite{ledig2017photo} applied a Generative Adversarial Network (GAN)~\cite{goodfellow2014generative} to SR, allowing better resolution of high-frequency details. 
These works focus on 2D planar data, which may not be ideal for 360-degree image super-resolution due to the distortions introduced in the projected representation. 
Our proposed model, however, operates directly on spherical signals so that we can avoid the distortion problem.

Focusing on optimizing 360-degree video streaming, 
Chen et al.~\cite{10.1145/3386290.3396929} focused to apply super-resolution on 360-degree video tiles. Their work mainly focused on the overall video streaming system rather than the super-resolution model implementation. This is different from our work that mainly focuses on the implementation of a novel spherical super-resolution model for 360-degree videos.

\section{Methodology}

\subsection{Focused icosahedral mesh} 
In this section, we first introduce our proposed focused icosahedral mesh. 
The icosahedral spherical mesh~\cite{baumgardner1985icosahedral} is a common discretization of the spherical surface. The mesh can be obtained by progressively sub-dividing each face of the unit icosahedron into four equal triangles. 

Operations on a full spherical mesh, refined to a granularity that can include all pixels from a planar representation of a 360-degree video frame, however, requires a significant amount of computation. 
In addition, operations on the full mesh cannot easily support operations on sub-areas of the spherical surface. 

\begin{figure}[t]
    \centering
    \includegraphics[width=0.7\columnwidth]{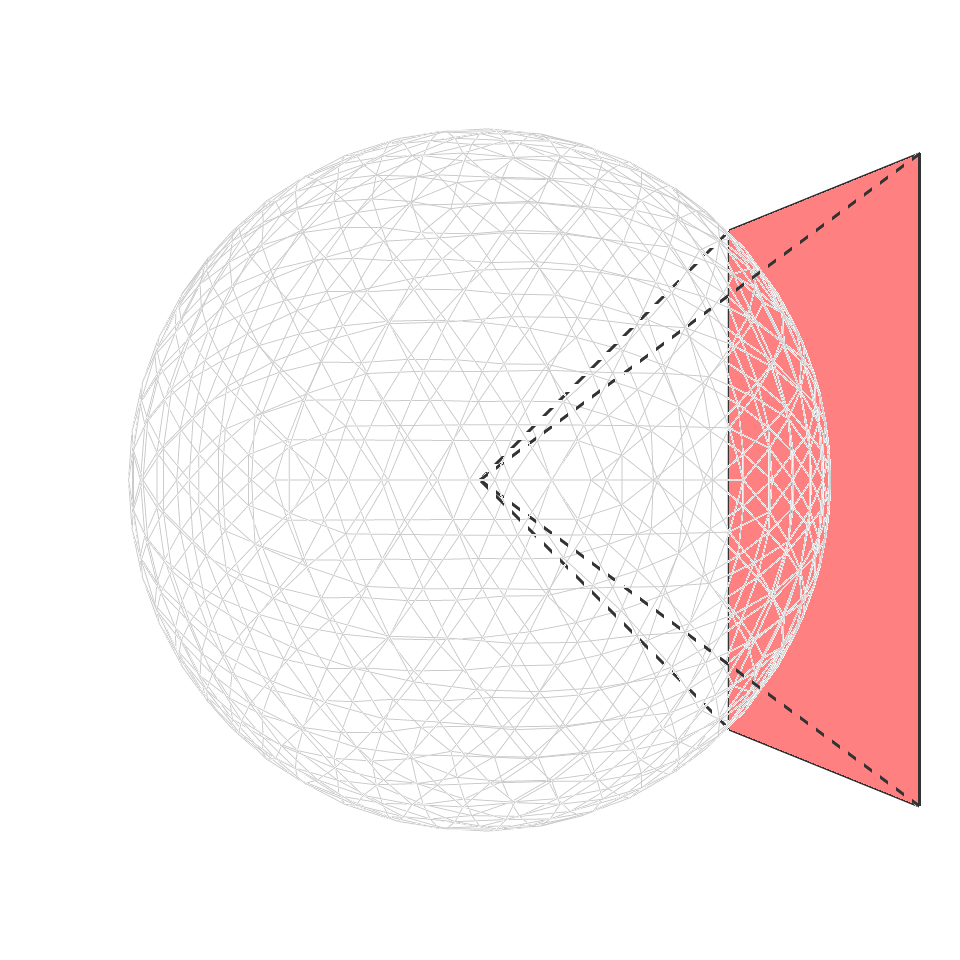}
    \caption{The user can only observe a sub-portion of the sphere at a time. For example, this figure shows a 80-degree by 80-degree field-of-view. }
    \label{fig:fov}
\end{figure}

Performing super-resolution on ``sub-areas'' of the spherical surface can be beneficial for real-world 360-degree applications. This is because human eyes as well as their viewing devices (e.g., the head-mounted display) have limited fields-of-view (FoV), usually represented as the angular extent of the field that can be observed.
For example, Figure \ref{fig:fov} represents a 80-degree by 80-degree FoV. 
To render the view shown in this figure, only part of the sphere is required. 

Such ``sub-areas'' would be useful in ``tiling'' schemes that can be used to support spatial-adaptive super-resolution over the 360-degree view.
That is, if only a small area on the sphere will be viewed by the user, we may only need to apply super-resolution to a sub-portion of the sphere instead of the full sphere.
As a result, performing super-resolution on the full icosahedral mesh may no longer be necessary as it requires more computation resources.

To support both faster operation and super-resolution on a sub-portion of the sphere, we propose a partial refinement scheme to generate ``Focused Icosahedral Mesh''.

\begin{figure*}[t]
\centering
\subfigure[\textbf{Full} Level-1 Mesh]{
	\includegraphics[height=.19\textwidth]{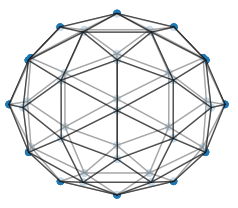}
}
\subfigure[\textbf{Focused} Level-2 Mesh]{
	\includegraphics[height=.19\textwidth]{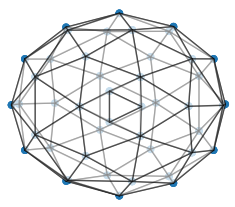}
}
\subfigure[\textbf{Focused} Level-3 Mesh]{
	\includegraphics[height=.19\textwidth]{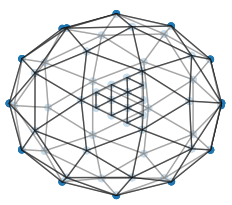}
}
\subfigure[\textbf{Focused} Level-5 Mesh]{
	\includegraphics[height=.19\textwidth]{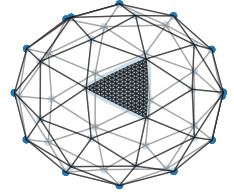}
}

\caption{Example of meshes in Level-1, Level-2, Level-3 and Level-5. To create ``Focused icosahedral meshes'', we select one face in the full Level-1 mesh and repeatedly refine triangles in this Level-1 face to obtain Focused Level-X meshes.}
\label{fig:mesh}

\end{figure*}

To generate a focused icosahedral mesh, we first create a Level-1 icosahedral mesh by refining each face on a unit icosahedron into 4 faces. In this way, the 20-face icosahedron is refined into a Level-1 icosahedral mesh with 80 faces.
An example full Level-1 mesh with 80 faces is shown in Figure \ref{fig:mesh}(a).

We then select one face out of the 80 faces of the Level-1 icosahedral mesh and only refine triangles located inside the selected Level-1 face. 

Specifically, in our focused mesh representation, we select the face of the Level-1 mesh that covers the position of $<$latitude=0, longitude=0$>$ on the sphere since very little distortion is introduced when pixels near this area are projected to the 2D plane.
Figure \ref{fig:mesh}(b) shows the Focused Level-2 mesh where the selected Level-1 face is refined into 4 smaller faces.
Figures \ref{fig:mesh}(c) and \ref{fig:mesh}
(d) show the Focused Level-3 and Focused Level-5 meshes, respectively. 

\subsubsection{Rotating content to the Focused Level-1 origin} 

Our model operates on a single focused icosahedral mesh, instead of operating on separate meshes for different Level-1 refined icosahedral faces. 
To allow our model to perform super-resolution for any area on the sphere, we need to map spherical pixel content that belongs to any arbitrary full Level-1 mesh face to the face that is selected to be refined.  
To do so, we pre-compute a rotation matrix $M \in R^{F \times V \times C} $, where $F$ represents the total number of faces in a full Level-1 mesh, which is $80$. $V$ is the number of vertices in a Level-1 face, as shown in figure\ref{fig:mesh}. 
A Level-1 mesh consists of $80$ triangles, each containing $3$ vertices.  
$C$ represents the number of dimensions of Euclidean coordinates in sphere, namely \textit{xyz}.

We denote the Level-1 face selected to be refined as face $F_0$. 
To rotate an arbitrary face $F_i, i\in(0,80)$ on the Level-1 mesh to the refined face $F_0$, we need to find a rotation matrix $M_i$ for face $F_i$ such that $F_i = M_i\cdot F_0$, where $F_i$ and $F_0$ are $3\times3$ matrices that represent the \textit{xyz} coordinates of three vertices of a triangle face.

Therefore, we can obtain $M_i$ as: $M_i = F_i\cdot F_0^{(-1)}$. 
We first rotate the vertices in the Focused Level-X Mesh with the rotation matrix $M$, and then compute a mapping from each pixel in the input planar representation (e.g., equirectangular image) to the rotated Focused Level-X vertex. 
In this way, we can represent all 80 different faces on the full Level-1 mesh through a single Focused Mesh file, which has the potential to save a significant amount of computation and storage resources and achieves better parameter efficiency.

\begin{figure}[!h]
\centering
\subfigure[This figure shows an equirectangular-projected 360-degree image. Magenta circles, b and c, in this figure mark areas corresponding to two different refined faces. (Original photo by Timothy Oldfield on Unsplash: \protect\url{https://unsplash.com/photo/luufnHoChRU})]{
\includegraphics[width=0.48\textwidth]{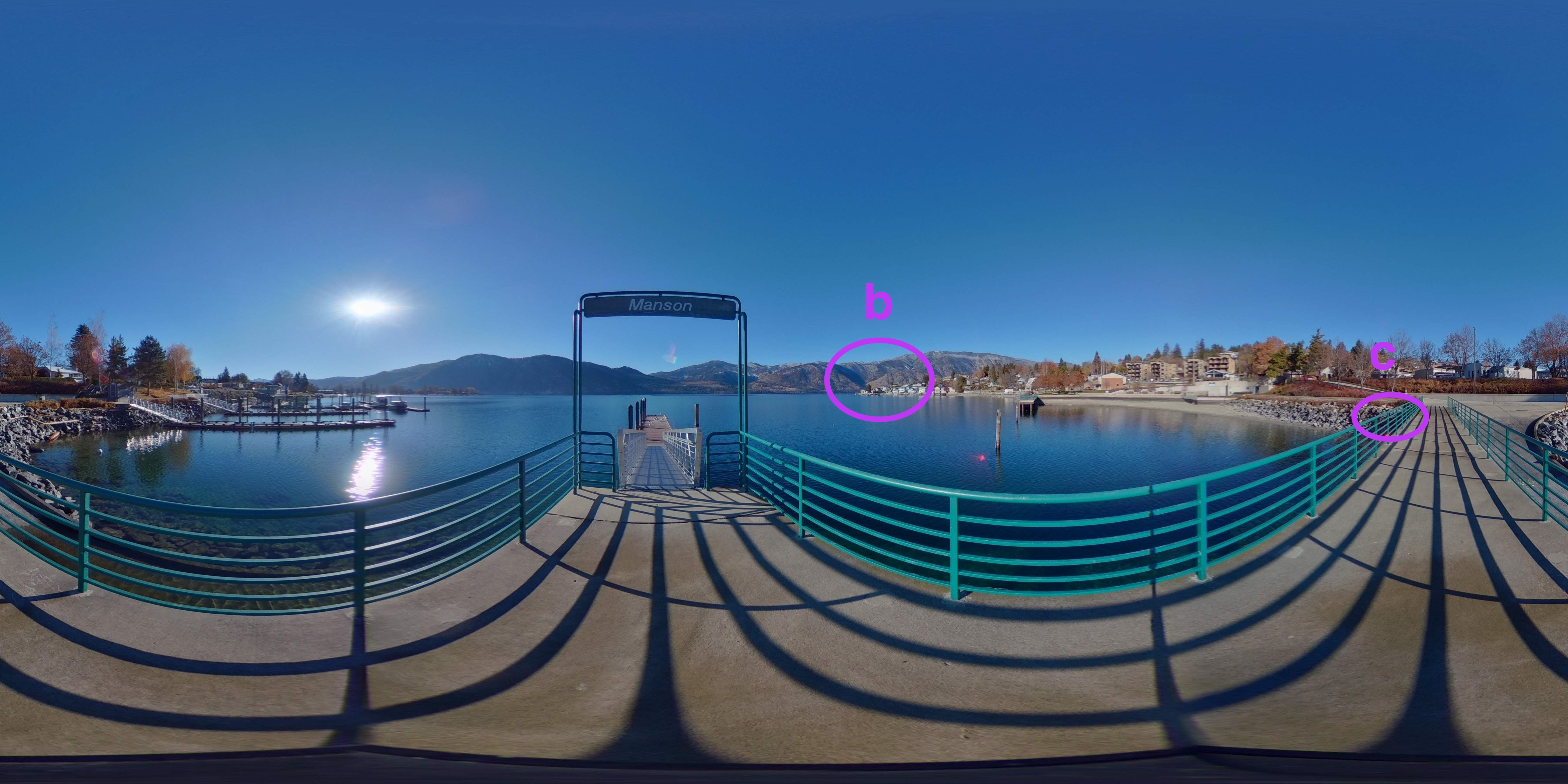}}
\subfigure[
The left-hand-side image displays the Focused Level-9 mesh visualized on an equirectangular image. 
The right-hand-side image displays a magnified view of the refined face. 
In both images, magenta points represent vertices in the full Level-1 mesh.]{
\includegraphics[width=0.33\textwidth]{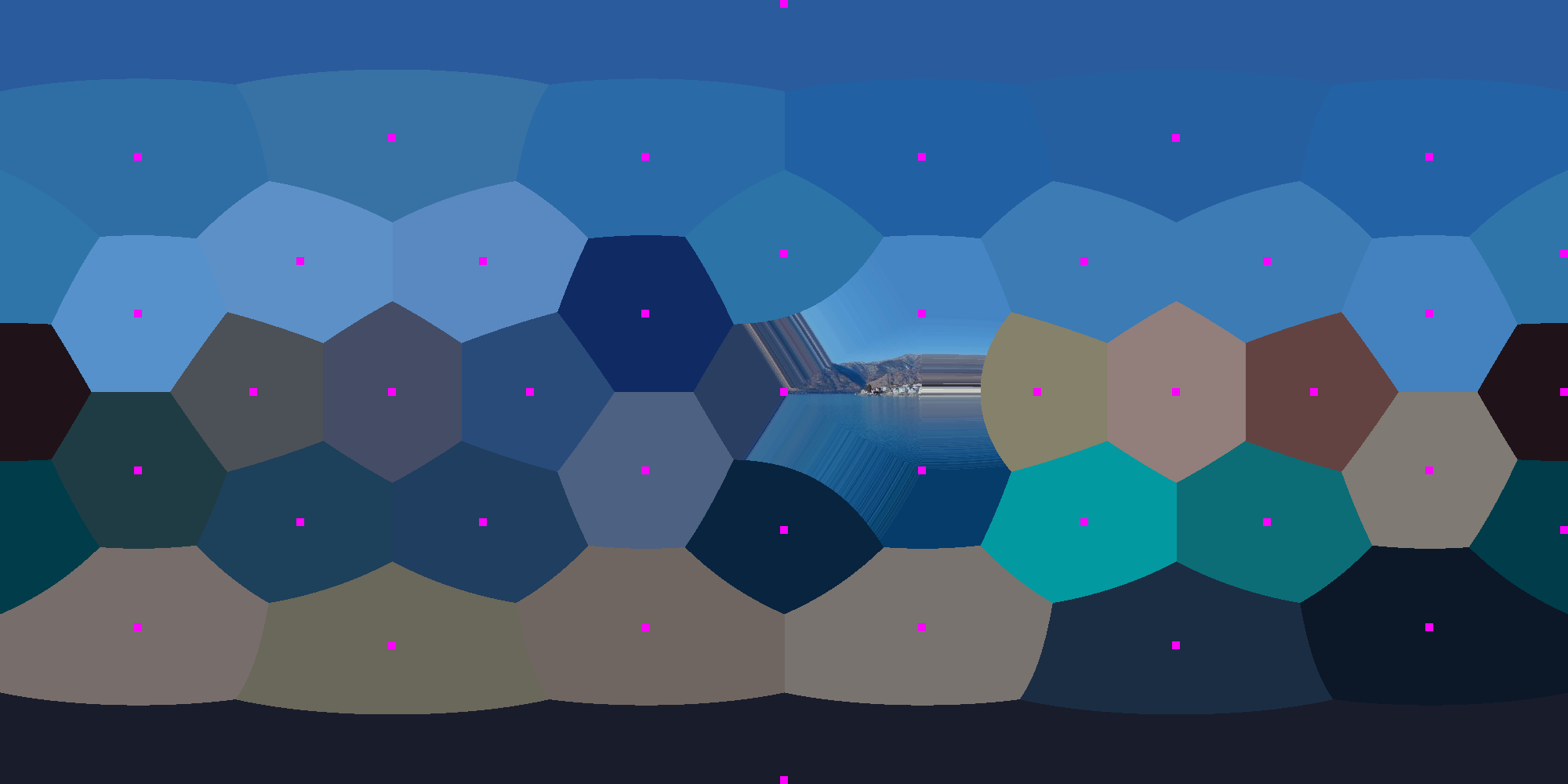}
\includegraphics[width=0.15\textwidth]{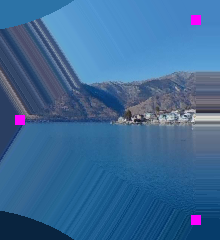}}
\subfigure[
This figure displays a different Level-1 icosahedral face rotated to the face refined in the Focused Mesh. Pixel values from the original image are attached to rotated vertices by inverting the rotation for positions of the mesh vertices then finding the nearest neighbor pixel of this rotated position.]{
\includegraphics[width=0.33\textwidth]{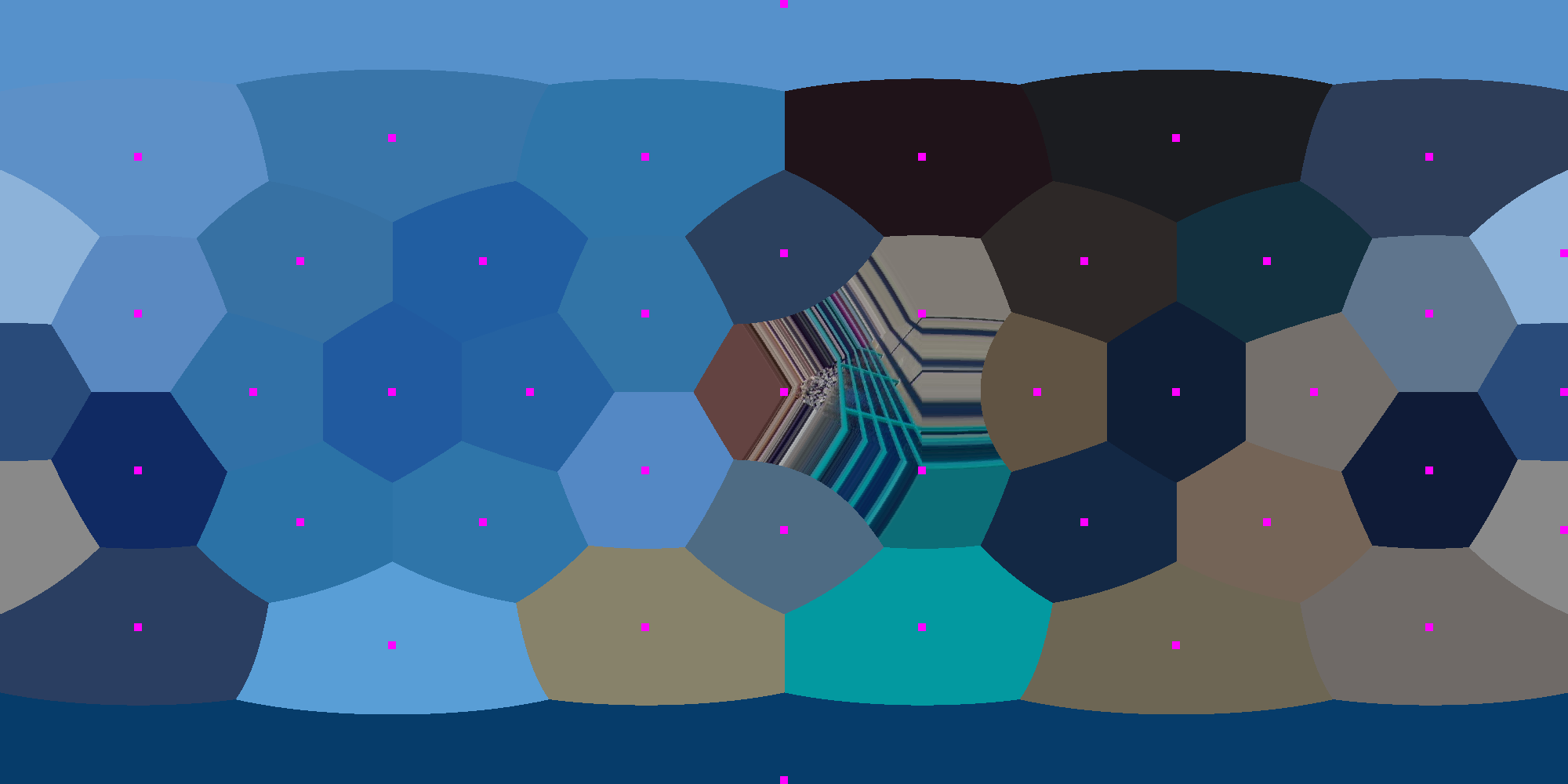}
\includegraphics[width=0.15\textwidth]{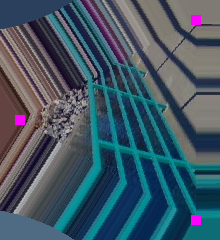}}
\caption{Visualizing the Focused Icosahedral mesh.}
\label{fig:visualize}
\end{figure}

Figure \ref{fig:visualize} visualizes how one focused icosahedral mesh can be used to represent all 80 different Level-1 faces. 
Figure \ref{fig:visualize}(a) shows an original equirectangular-projected 360-degree image. 
In this image, we highlight two areas marked by magenta circles.
In Figure \ref{fig:visualize}(b), the left-hand-side image shows the Focused Level-9 mesh visualized on an equirectangular image. 
Magenta points in this figure represent vertices in the full Level-1 mesh. 
There are 42 vertices in the full Level-1 mesh.
The right-hand-side image in Figure \ref{fig:visualize}(b) magnifies the refined face in the Focused icosahedral mesh to show details. 
We can see that content in this face are in the same position as in the original equirectangular-projected image. 

Figure \ref{fig:visualize}(c) shows the resulting visualization when we rotate a different Level-1 face to the refined face. The image on the right magnifies the refined face to show details. 

\subsubsection{Mesh sizes}

Table \ref{tbl:mesh} shows the number of vertices in both Full and Focused icosahedral meshes in different levels of refinement. 
A Full Level-9 mesh has more than 2.6 million vertices and requires more than 1.9 GB space for storage. 
On the other hand, a Focused Level-9 mesh has only about 33K vertices, requiring only about 31 MB storage space. 

\begin{table*}[t]
\centering
\caption{Number of vertices in \textbf{Full} icosahedral mesh, \textbf{Focused} icosahedral mesh, and their roughly-equivalent 2D planar resolution in the equirectangular projection.}
\vspace{1em}
\label{tbl:mesh}
\begin{tabular}{@{}lrrrr@{}}
\toprule
Level & Level-6      & Level-7      & Level-8       & Level-9        \\ \midrule
Full    & 40,962   & 163,842  & 655,362   & 2,621,442   \\
Focused & 600     & 2,184    & 8,424     & 33,192     \\ \midrule
2D planar    & 360x180 & 720x360 & 1440x720 & 2880x1440 \\ \bottomrule
\end{tabular}
\vspace{-1em}
\end{table*}

We know that the area of a unit spherical surface is $4\pi$.
A frame generated through the equirectangular projection covers a corresponding area of $2\pi \times \pi = 2\pi^2$. 
Suppose there are $N_x$ vertices in the Full Level-X mesh, given that vertices on the icosahedral mesh are roughly uniformly distributed on the sphere, we can estimate the equivalent 2D equirectangular-projected frame resolution as follows: $W = \sqrt{N_x\times\pi}$, $H = W / 2$, where $W$ and $H$ are the width and height of the equirectangular projection, respectively.
The results are listed in Table \ref{tbl:mesh}.
We find that Level-6 mesh is roughly equivalent to the 2D equirectangular projection in 360x180 resolution, and that Level-9 mesh is roughly equivalent to the 2D equirectangular projection in 2880x1440 resolution.

\subsection{MeshConv Transpose} 
UGSCNN\cite{jiang2019spherical} also proposed MeshConv Transpose operation in their UNet architecture. MeshConv Transpose takes level-$i$ mesh for input and outputs a level-$i+1$ mesh, which can be described as follows:

\begin{align*}
    M_{i+1} = \text{MeshConv}(\text{Padding}(M_i))
\end{align*}

where $P$ represents zero padding, $M_{i+1}$ and $M_i$ are level-$i+1$ mesh and level-$i$ mesh, respectively. In general, MeshConv Transpose simply padding $0$s on new vertexes in level-$i+1$ mesh, then apply MeshConv on the new zero-padding level-$i+1$ mesh. However, it's easy to implement but inefficient. 

\subsection{VertexShuffle}

Motivated by PixelShuffle~\cite{shi2016real} commonly used in 2D super-resolution models, we proposed VertexShuffle in our spherical super-resolution model, which can be described as follows:

\begin{align*}
& M_{i+1} = \text{VertexShuffle}(M_i) \\
& M_i = {M_{i0}, M_{i1}, M_{i2}, M_{i3}} \\
& N'_i = (M_{ij} + M_{i(j+1)}) / 2 \\
& N_i = unique(N'_i) \\
& \text{VertexShuffle}(M_i) = concat(M_{i0}, N_i)
\end{align*}

The input of our basic VertexShuffle operation can be represented as $M_i \in R^{F \times V_i}$, where $F$ is the feature dimension in Level-$i$, and $V_i$ represents the number of vertices of Level-$i$ mesh. 
The output is $M_{i+1} \in R^{F' \times V_{i+1}}$, where $F'$ is the feature dimension in Level-$i+1$, which is $F/4$ in our work, and $V_{i+1}$ represents the number of vertices of the Level-$i+1$ mesh.

We firstly split $M_i$ into four parts $\{M_{i0}, M_{i1}, M_{i2}, M_{i3}\}$ on feature map dimension, where $M_{ij} \in R^{F' \times V_i},j=0,1,2,3$, $F'$ here is $F/4$. We keep $M_{i0}$ as our Level-$i$ mesh, which will be used later. $M_{i1}, M_{i2}, M_{i3}$ are used to refined vertices in Level-$i+1$ mesh. 

As we introduced before, a spherical mesh can be obtained by progressively sub-dividing each face of the unit icosahedron into four equal triangles. Here, we treat a single triangle face as a sequence of vertices, $v_0, v_1, v_2 $ and a sequence of edges $(v_0,v_1), (v_1,v_2), (v_2,v_0)$. The refinement process can be regarded as progressively construct midpoint vertex on associated edges, and new edges in Level-$i+1$ are created between each pair of midpoint vertices, thus a single face in Level-$i$ are refined into four new faces in Level-$i+1$.

To fully make use of feature maps in Level-$i$, we use $M_{i1}, M_{i2}, M_{i3}$ to refine vertices in Level-$i+1$ mesh. Specifically, we use $M_{i1}$ to calculate midpoint between $(v_0,v_1)$, $M_{i2}$ to calculate midpoint between $(v_1,v_2)$, and $M_{i3}$ to calculate midpoint between $(v_2,v_0)$. Midpoint vertex values are constructed by averaging the values associated with the original two vertices on a edge, which can be described as follows:

\begin{align*}
N'_{i0} & =  (M_{i1}(v_0) + M_{i1}(v_1)) / 2 \\
N'_{i1} & =  (M_{i2}(v_1) + M_{i2}(v_2)) / 2 \\
N'_{i2} & =  (M_{i3}(v_2) + M_{i3}(v_0)) / 2
\end{align*}

Thus, we can get a set of midpoint vertices $N'_i$, which are new vertices generated in Level-$i+1$ mesh. However, there exits redundant midpoints due to the shared edges that may be calculated twice. We have to performs deduplication on the set of midpoint vertices. There are plenty of ways to select midpoint between the two calculated midpoint, such as, \texttt{max}, \texttt{min}, \texttt{average}, \texttt{weighted average}. In our paper, we simply select the first instance of a midpoint, which achieves best results in our experiments. 

Then, we have a set of unique midpoint vertices that used to refine the next level mesh $N_i \in R^{F' \times A_i}$, where $A_i = V_{i+1} - V_i$. 

Finally, we concatenate partial of feature map in Level-$i$ $M_{i0}$ with the new calculated midpoint vertices $N_i$ to formulate our Level-$i+1$ mesh.

Compared to MeshConv Transpose, we do not have extra learnable parameters. In other word, the implementation of VertexShuffle is not only more parameters efficient, but also achieves significantly better performance.

\begin{figure*}[t]
\centering
\includegraphics[scale=0.56]{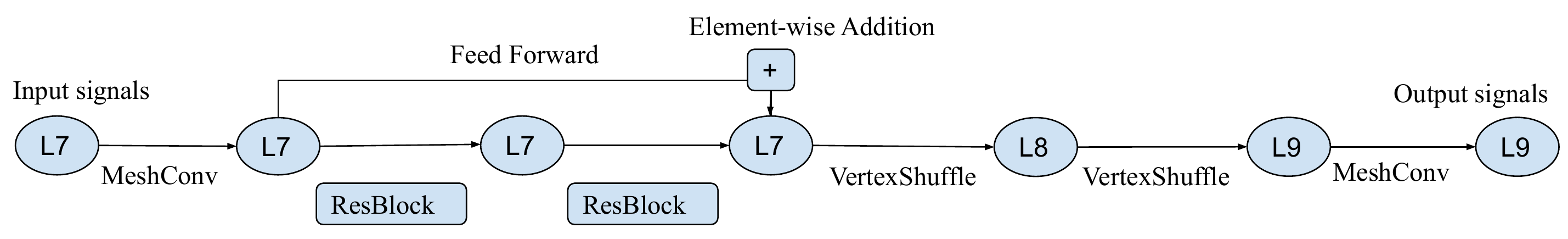}
\caption{This figure shows the architecture of our proposed Spherical Super-Resolution (\textbf{SSR}) model that uses MeshConv and VertexShuffle operations. Here, L7 represents the input Level-7 mesh, and L9 represents the output Level-9 mesh. Our model starts with a MeshConv layer followed by 2 ResBlocks and 2 VertexShuffle layers, it then ends with a final MeshConv layer.}
\label{fig:ssr}
\end{figure*}

\begin{figure}[t]
\centering
\includegraphics[scale=0.42]{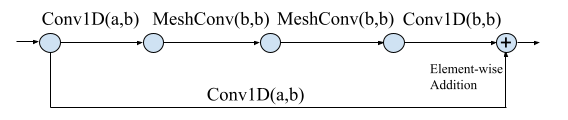}
\caption{The adapted ResBlock used in our model.}
\label{fig:res}
\end{figure}

\subsection{Model architecture} 

We apply our Focused icosahedral mesh and VertexShuffle in super resolution.
The architecture of our model is shown in Figure \ref{fig:ssr}. 
In this figure, we show the input of our model as a Level-7 Focused icosahedral mesh, it first goes through a MeshConv layer with Batch Normalization~\cite{10.5555/3045118.3045167} followed by a ReLU~\cite{agarap2018learning} activation function. Then, we use two adapted Residual Blocks~\cite{he2016deep} to further extract features, which we will explain in depth later. We further concatenate the output of the first MeshConv and the output from the two ResBlocks by element-wise addition. After that, we apply two VertexShuffle operations to upscale the features. Finally, our model ends up with a MeshConv layer. Thus, a Level-9 Focused icosahedral mesh is generated.

\noindent
\textbf{Adapted Residual Block.}
We adapt a regular residual block by adding two MeshConv layers in the residual block. Other settings are similar to the regular residual block~\cite{he2016deep}. 

\noindent
\textbf{MeshConv.}
The MeshConv operation introduced by Jiang~\cite{jiang2019spherical} et al. is performed by taking a linear combination of linear operator values computed on a set of input mesh vertex values. MeshConv can be formulated as follows:

$$
\text{MeshConv}(F;\; \theta)
 = \theta_0 IF + \theta_1 \nabla_{lat}F + \theta_2\nabla_{lng}F + \theta_3\nabla^2F
$$

where $I$ represents for the identity, which can be regarded as the $0th$ order differential, same as $\nabla_{00}$. $\nabla_x$ and $\nabla_y$ are derivatives in two orthogonal spatial
dimensions respectively, which can be viewed as the $1st$ order differential. $\nabla_2$ stands for the Laplacian operator, which can be regarded as the $2nd$ order differential.

At a high-level, these linear operators can be viewed as computing a set of local information near each vertex of the mesh. The standard $3 \times 3$ cross correlation operation can be viewed as a set of nine linear operators. Each of the linear operators returning a value of either the pixel itself or an adjacent pixel.
Compared to the $3 \times 3$ convolution, it is clear that the set of four linear operators used by MeshConv are less expressive. They not only extract less information per pixel, but this information also can drop information about a vertex's surrounding. For example, the gradient operation on the mesh computes a 3-dimensional average of either six or seven values. Another degree-of-freedom is dropped from the gradient when taking only the east-west and north-south components of the gradient.
We hypothesize that some of the information excluded from the linear operator computations could be useful for the super-resolution task. To attempt to mitigate this information loss, rather than including single MeshConv ops in our network architecture, we include pairs of composed MeshConv ops. These paired operations aggregate more local information around a vertex before the non-linearity is applied, allowing the network to capture more-useful characteristics needed for the super-resolution task.

\subsection{Loss function}

Similar to general super-resolution tasks, our goal is minimizing the loss between the reconstructed images $Y_i$ and the corresponding ground truth high-resolution images ${H_i}$. Given a set of high-resolution images ${H_i}$ and their corresponding low-resolution images ${X_i}$, we represent the loss as follows:

$$
\text{MSE} = \frac{1}{N}\sum_{i=1}^N(F(X_i)-Y_i)^2
$$

$$
\text{L} = 10\times \log_{10}(\text{MSE})
$$

where $N$ is the number of training samples. 
Here, we adapt mean square error as our loss function. It can be regarded as the negative peak signal-to-noise (PSNR) value, which is more straightforward in our task. 

\begin{table*}[!t]
\centering
\caption{PSNR (dB) results for small and large dataset.}
\label{tbl:psnr}
\begin{tabular}{@{}l|cc|c@{}}
\toprule
Model   &  Small   & Large  & Average  \\ \midrule
Spherical: MeshConv with transposed MeshConv    &18.52 &16.57 &17.54 \\
Spherical: MeshConv with VertexShuffle (\textbf{SSR})  &31.44 &34.13 &32.79  \\
\bottomrule
\end{tabular}
\end{table*}

\begin{table*}[!t]
\centering
\vspace{1em}
\caption{Comparison of total number of model parameters, and per-frame inference time.}
\label{tbl:size}
\begin{tabular}{@{}l|c|c@{}}
\toprule
Model   &  Total \# of Parameters  & Per-image/frame Inference Time \\ \midrule
Spherical: MeshConv with transposed MeshConv &1001225   & 5883 ms\\
Spherical: MeshConv with VertexShuffle (\textbf{SSR}) & 734905 & 578 ms\\
 \bottomrule
\end{tabular}
\end{table*}

\section{Experiments}
\subsection{Dataset}
Due to the lack of official spherical super-resolution datasets, we collect two publicly-available 360-degree video datasets: the 360-Degree Video Head Movement Dataset~\cite{corbillon2017360} and VR User Behavior Dataset~\cite{Chenglei:2017:vrdataset} to generate a spherical super resolution dataset with high quality. 
The 360-Degree Video Head Movement Dataset~\cite{corbillon2017360} contains 5 videos in 4K quality, and the VR User Behavior Dataset contains 5 videos with $2560 \times 1440$ resolution. 
We use FFmpeg~\cite{ffmpeg} to extract the key frames of each video in the dataset.

We first construct a small dataset with the 360-Degree Video Head Movement Dataset~\cite{corbillon2017360}. This small dataset contains 345 images of $2880 \times 1440$ resolution. We randomly split the dataset with $80\%$ training set and $20\%$ test set. The training set provide roughly 21,440 training items, and the test set provide 6,160 testing items. We evaluate our model with the upscaling factor of 4, that is, 16x super-resolution.

We also generate a larger dataset with both the 360-Degree Video Head Movement Dataset~\cite{corbillon2017360} and the VR User Behavior Dataset~\cite{Chenglei:2017:vrdataset}.
The large dataset contains 1,532 images in total with $2560 \times 1440$ resolution. As with the small dataset, we split the dataset with $80\%$ training set and $20\%$ test set. The training set provide roughly 95,360 training items, and the test set provide 27,200 testing items. We evaluate our model with the same upscaling factor of 4.

\subsection{Implementation details}
With our generated Focused Icosahedral mesh, we map the 2D equirectangular-projected frame to the partial mesh sphere. In both the small and large datasets, the input data is in Level-7, which is roughly equivalent to a 2D equirectangular-projected frame in $720 \times 360$ and $640 \times 360$ resolution. The output data is in Level-9, which is roughly equivalent to $2880 \times 1440$ and $2560 \times 1440$ equirectangular-projected frame, respectively. The upscaling factor in our experiment set up is $\times4$. That is, the number of output vertices is 16x the number of input vertices.

In our experiments, we train our model with 50 epochs with batch size of 64. we set learning rate as 0.01 and use Adam~\cite{kingma2014adam} as our optimizer. 
We use the PSNR as the performance metric to evaluate our models.

\subsection{Comparison with MeshConv Transpose}
We compare our spherical super-resolution (SSR) model that uses the VertexShuffle operation with a baseline model that uses the MeshConv Transposed proposed in the original UGSCNN paper ~\cite{jiang2019spherical}. 
We conduct experiments on both of the small and large datasets with same configuration, i.e., using Focused Icosahedral mesh with an upscaling factor of $\times 4$.

Table \ref{tbl:psnr} shows the performance of two models on both datasets. As we can see, our model achieves significantly higher performance than the baseline MeshConv transpose model. We also found that with the increasing volume of dataset, the performance of our model gets better. Our model can achieve the PSNR result on larger dataset of 34.13 dB, 31.44 dB on small on, and 32.79 dB on average, while the baseline method performs not so good in spherical super resolution task, it can only achieve 17.54 dB on average. 

\begin{figure*}[!t]
    \centering
    \includegraphics[width=2.05\columnwidth]{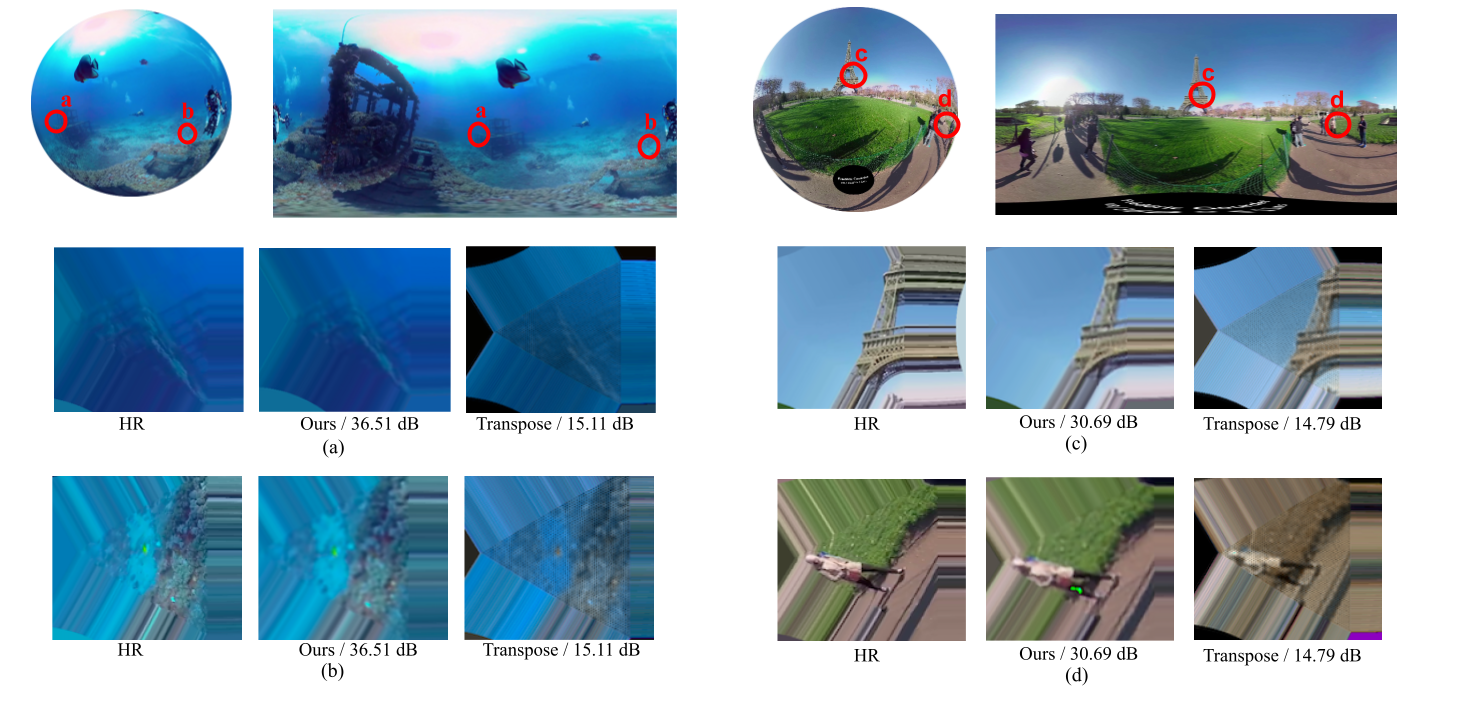}
    \caption{Visualization of results of our proposed SSR model (represented as ``Ours'') and the MeshConv transpose model (represented as ``Transpose''). ``HR'' represents the high resolution version of the data, i.e., the groundtruth. }
    \label{fig:vis}
\end{figure*}

\subsection{Inference time and parameters}
For fair comparison of model inference time, we fixed the batch sizes of both models to 16. 
We first compute the average inference time for each batch, then divide it by the batch size, and finally, we multiple the number of data items for a full frame, which is $80$ here. 
This allows us to roughly obtain the per-frame inference time. 
We collected our results via experiments on a Tesla P100 GPU.

Table \ref{tbl:size} compares the inference time and total number of parameters of our proposed model with the baseline model that uses MeshConv transpose.

Our model can save roughly 20\% parameters compared to the baseline model, which improves the efficiency significantly. 
In addition, MeshConv with transposed MeshConv requires nearly 6 seconds to process a full image/frame.
With our proposed VertexShuffle operation, our SSR model can achieve more than 10x acceleration in processing a full image/frame, which is significantly faster than the spherical baseline model with transposed MeshConv operations. 

In addition, given that a user can only watch a sub-portion of the spherical image/frame at a time, there is no need to perform super-resolution for all 80 faces of a frame at the same time. 
This indicates that our SSR model can be used in real-world applications with even faster per-image/frame processing time.

\subsection{Quantitative results}
Overall, our model outperforms the baseline MeshConv-transpose-based model in all of PSNR results, inference time, and the total number of parameters. Since we are the first to directly apply spherical convolutional neural network on super-resolution task, we have no benchmark to compare with. Simply comparing the PSNR results with the 2D super-resolution task would be unfair due to the different data format convolution method. Hence, we only compare our method with the MeshConv transpose model, which is a fair comparison to show our contributions. 

\subsection{Qualitative results}
Figure \ref{fig:vis} visualizes the results of our proposed SSR model and the MeshConv transpose model. 
We use two images with significantly different PSNR results as examples.\footnote{These images are from the following two videos in the 360-degree video head movement dataset~\cite{corbillon2017360}: \url{https://www.youtube.com/watch?v=2OzlksZBTiA} and \url{https://www.youtube.com/watch?v=sJxiPiAaB4k}} 
Here, we first show the full image frame via two formats, in both the spherical domain and the 2D planar domain. Moreover, in figures (a) to (d), we select two focused icosahedral meshes from different locations on a sphere to demonstrate the efficiency of directly applying spherical super-resolution to spherical data.

\subsection{Discussion}

The computer vision community has made tremendous progress in 2D super-resolution while there is little precedent effort in directly applying spherical super-resolution. There are significant challenges in directly performing super-resolution on spherical signals, e.g., how to perform convolution operations in 3D space, how to perform deconvolution operations in 3D space, how to perform PixelShuffle in 3D space with other spherical convolution method, etc. 
In this paper, we provide a straightforward approach to directly apply 3D convolution to spherical signals and can achieve good results, which shows great potential in the 3D super-resolution area.
In addition, we show that it is feasible to directly apply spherical super-resolution on spherical signals, which can avoid issues in applying 2D super-resolution in 3D space, such as distortion, oversampled pixels, etc.
We believe there are a great number of interesting directions to exploit ahead on directly apply super resolution on spherical signals.

\section{Conclusion}
In this paper, we proposed a memory- and bandwidth-efficient representation of the spherical mesh -- the Focused Icosahedral Mesh, which is more flexible than full meshes and saves a significant amount of computation resources, and a novel VertexShuffle operation to further improve the performance compared to the baseline model that uses the MeshConv Transpose operation. To illustrate our proposed idea, we present a spherical super resolution model and show great capacity and potential to apply the traditional 2D computer vision tasks on spherical signals. For evaluation, we create a new high-resolution spherical super resolution dataset by extracting key frames from a set of collected 360-degree videos. Experiments on the dataset show that our proposed model is superior to the baseline model in performing spherical super-resolution tasks with remarkable efficiency.

\bibliographystyle{bib}
\bibliography{main}

\end{document}